%% file: chapter.tex
\documentclass[journal = jctcce]{achemso}
\usepackage{amsmath}
\usepackage{amssymb}
\usepackage{dcolumn} 
\usepackage{setspace}
\usepackage{array}
\usepackage{graphicx}
\usepackage{booktabs}
\usepackage{multirow}
\usepackage{psfrag}
\usepackage{listings}

\nocite{*} 
\input{macros}

\def\sgtg{{G(T)$_N$G}}
\def\sgag{{G(A)$_N$G}}

\def\brp{{\mathbf{r}^{\prime}}}

\def\br{{\mathbf{r}}}

\def\bR{{\mathbf{R}}}
\def\bS{{\mathbf{S}}}

\def\d{{\mathrm{d}}}
\def\rhor{{\rho({\bf r})}}

\def\rhoi{{\rho_I}}

\def\rhoir{{\rho_I({\bf r})}}

\graphicspath{{./figures/}}

\author{Pablo Ramos}
\author{Marc Mankarious}
\author{Michele Pavanello}
\email{m.pavanello@rutgers.edu}
\affiliation{Department of Chemistry, Rutgers University, Newark, NJ 07102, USA}
\title{A critical look at methods for calculating charge transfer couplings fast and accurately}

\begin{document}
\maketitle

\begin{abstract}
We present here a short and subjective review of methods for calculating charge transfer couplings. Although we mostly focus on Density Functional Theory, we discuss a small subset of semiempirical methods as well as the adiabatic-to-diabatic transformation methods typically coupled with wavefunction-based electronic structure calculations. In this work, we will present the reader with a critical assessment of the regimes that can be modelled by the various methods -- their strengths and weaknesses. 
In order to give a feeling about the practical aspects of the calculations, we also provide the reader with a practical protocol for running coupling calculations with the recently developed FDE-ET method.
\end{abstract}
\newpage
\tableofcontents

\newpage
\section{Introduction}
Charge transfer (CT) between molecular species play vital roles in processes that occur in biology such as protein communication\cite{grey1996,kawa2006,Larsson1983,Farid1993,Gray2005}, respiratory systems in the mitochondria\cite{Gennis1994}, oxidative damage on DNA\cite{bixo1999,gies2004,gies2002}, photosynthetic cycles \cite{Stowell02051997,Balabin06102000}, as well as  in materials science conduction in organic semiconductors\cite{Troisi2011,mcma2011}. In order to achieve an accurate modeling of these processes in the simulations, one needs to include several levels of complexity, which in most instances lead to considering model systems featuring hundreds of atoms and an even larger number of electrons. The large system sizes preclude the use of high-level wavefunction-based quantum-chemical methods. For this reason, researchers worldwide have invested a great deal of effort in developing approximate, fast, yet still accurate methods for describing CT reactions. Methods based on Density-Functional Theory have in recent years become competitive in regards to the accuracy while still maintaining a generally low computational cost. 

Marcus theory\cite{marcus1956,marcus1985} is perhaps the most applicable theory for modeling a CT process. This theory was originally derived under three main approximations. First, a CT event is thought of in terms of a two-dimensional basis set (donor and acceptor). The interaction matrix element of the Hamiltonian is the central quantity in determining the probability of a transition in populations from the basis function representing the donor state to one representing the acceptor state. This interaction is the electronic coupling $V_{DA}$ of the two electronic states involved in the CT reaction\cite{landau1932,zene1932}. Second, it relies on the Condon approximation\cite{Condon1298,Franck1926}, in which the electronic coupling is considered to be independent of the nuclear motion when the transfer occurs. Third, reactants and products are modeled as being enclosed by spheres on which the polarization of the solvent is represented as a dielectric continuum \cite{Ladanyi1993,Ingram2003,Maroncelli1993,marcus1985}. Marcus theory can be summarized as\cite{nitzan_book}:

\begin{equation}
\label{marcus}
k_{CT} = \frac{2\pi}{\hbar} |V_{DA}|^2 \frac{e^{-\frac{(\Delta G + \lambda)^2}{4\lambda K_BT}}}{\sqrt{4\pi\lambda K_B T}}.
\end{equation}
where $\lambda$ is the reorganization energy, and $V_{DA}$ is the electronic coupling.

States that most resemble the initial and final states of electron transfer have been often referred to as ``diabatic states'' \cite{subo2009, vanv2010a} and their corresponding wavefunctions ``diabats''.
Although it is known that diabatic states have a formal definition \cite{london1932a,mead1982}, it was shown \cite{pava2011a} that charge-localized states satisfy the requirements for diabatic states for condensed phase electron transfer reactions.

Several approaches are available in the literature to generate and evaluate Hamiltonian matrix elements with wavefunctions of charge-localized, diabatic states. They differ in the level of theory used in the calculation and in the way localized electronic structures are created\cite{london1932a,cave1997,newton1991,wars1980,marcus1985,vanv2010a}. When wavefunction-based quantum-chemical methods are employed, the framework of the generalized Mulliken-Hush method (GMH)\cite{cave1997,mull1952,Hush1968,Cave1996}, is particularly successful. So far, it has been used in conjunction with accurate electronic structure methods for small and medium sized systems\cite{voityuk2006a,voityuk2007a,Kubas2014}. As an alternative to GMH and other derived methods \cite{fate2013,subo2008}, additional methods have been explored for their applicability in larger systems such as constrained density functional method (CDFT)\cite{kadu2012,wu2006,vanv2010a,Kubas2014}, and fragmentation approaches\cite{voityuk2002,hsu2008,migliore2011a,migliore2006,migliore2009c,cem2009}, which also include the frozen density embedding (FDE) method\cite{pava2011b,pava2013a}. 

So far, we have mentioned methods that produce all-electron diabatic wavefunctions and corresponding Hamiltonian matrix elements. There are two other classes of methods which simplify the quantum problem by focusing on the wavefunction of the transferred charge: such as methods making use of the frozen core approximation Fragment Orbital methods (FO), and methods that assume the charge to be localized on single atomic orbitals \cite{Beratan31051991}. In this work, we will also treat these computationally low-cost methods.

As our group is involved in the development of the Frozen Density Embedding (FDE) formulation of subsystem DFT, this chapter will pay particular attention to the FDE methodology. We believe FDE to be a very promising method capable of achieving a good description of the electronic coupling in CT reactions, while maintaining a low computational complexity.

This chapter is divided in two parts: the first part is devoted to the FDE method as well as other DFT-based alternatives. The second part covers more accurate methods (wavefunction-based). In each of the two parts, we discuss the numerical stability and accuracy of the methods in the generation of diabatic states with the overarching goal of obtaining reliable electronic couplings with a contained computational effort. 

We will start with a description of FDE and its ability to generate diabats and to compute Hamiltonian matrix elements -- the FDE-ET method (ET stands for Electron Transfer). In the subsequent section, we will present specific examples of FDE-ET computations to provide the reader with a comprehensive view of the performance and applicability of FDE-ET. After FDE has been treated, four additional methods to generate diabatic states are presented in order of accuracy: CDFT, FODFT, AOM, and Pathways. In order to output a comprehensive presentation, we also describe those methods in which wavefunctions methods can be used, in particular GMH and other adiabatic-to-diabatic diabatization methods. Finally, we provide the reader with a ``protocol'' for running FDE-ET calculations with the only available implementation of the method in the Amsterdam Density Functional software\cite{adf}. In closing, we outline our concluding remarks and our vision of what the future holds for the field of computational chemistry applyed to electron transfer. 
\section{DFT Based Methods}

\subsection{The Frozen Density Embedding formalism}
\label{FDEt}
The frozen density-embedding (FDE) formalism \cite{jaco2013} developed by Wesolowski and Warshel\cite{weso1993,weso2006,jaco2013} has been applied to a plethora of chemical problems, for instance, solvent effects on different types of spectroscopy\cite{neug2005b,neug2010a,HD2014}, magnetic properties \cite{jaco2006b,bulo2008,neug2005f,kevo2013,weso1999}, excited states\cite{neug2010,casi2004,pava2013b,neug2005b,garc2006},  charge transfer states \cite{ramo2014,pava2013a,solo2014}. Computationally, FDE is available for molecular systems  in ADF \cite{adf,jaco2008b}, Dalton \cite{dalton1_2,hofn2012a}, Q-Chem \cite{good2010,qchem}, and Turbomole \cite{Turbomoleb,ahlr1989,lari2010} packages, as well as for molecular periodic systems in CP2K \cite{cp2k,iann2006} and fully periodic systems (although in different flavors) in CASTEP \cite{CASTEP,laha2007}, Quantum Espresso \cite{fdeinqe,qe,genova2014}, and Abinit \cite{ABINIT,govi1998}. 

FDE prescribes that the total electron density should be expressed as the sum of subsystem electron densities\cite{weso1993,sen1986,cort1991,kolos1978,gord1972}, this is based on the idea that a molecular system can be more easily approached if it is subdivided into many smaller subsystems. Namely,
\begin{equation}
 \label{fdeden}
 \rho_{tot}(\br)=\sum_{I=1}^{\# of subsystems} \rho_I(\br).
\end{equation}
As in regular DFT calculations, the electron density of each subsystem is computed by solving selfconsistently a Kohn--Sham (KS) like equation per subsystem. These KS like equations read as: 
\begin{equation}
 \label{ksequ}
 \left[\frac{-\nabla^2}{2}+\upsilon^{I}_{KS}(\br)+\upsilon^{I}_{emb}(\br)\right]\phi_{(i)I}(\br)=\epsilon_{(i)I}(\br)\phi_{(i)I}(\br).
\end{equation}
Where $\phi_{(i)I}(\br)$, $\epsilon_{(i)I}$ are the molecular orbitals and orbital energies of subsystem $I$. In \eqn{ksequ} we have augmented the Kohn--Sham single particle Hamiltonian by an embedding potential, $\upsilon^{I}_{emb}$, in which are encoded the interactions with the other subsystems. In the following, $\upsilon^{I}_{emb}(\br)$ is the embedding potential acting on subsystem $I$:
\begin{align}
 \label{embpot}
 \nonumber
 \upsilon^{I}_{emb}(\br)=&\sum^{N_S}_{J\neq I}\left[\int \frac{\rho_J(\brp)}{|\br-\brp|}d\brp-\sum_{\alpha\in J} \frac{Z_{\alpha}}{|\br-\bR_{\alpha}|}\right]+ \\
&+\frac{\delta T_{\rm s}[\rho]}{\delta\rhor}-\frac{\delta T_{\rm s}[\rhoi]}{\delta\rhoir}+\frac{\delta E_{\rm xc}[\rho]}{\delta\rhor}-\frac{\delta E_{\rm xc}[\rhoi]}{\delta\rhoir}.
\end{align}
In the above, $T_{\rm s}$, $E_{\rm xc}$ and $Z_\alpha$ are kinetic and exchange--correlation energy functionals, and the nuclear charge, respectively, and $N_S$ is the total number of subsystems considered. In practical FDE calculations, the kinetic energy is calculated in terms of orbital free semilocal functionals. This approximation is ultimately the biggest difference between an FDE and a full KS-DFT calculation of the supersystem\cite{gotz2009,weso1996,lude2003,Lude2002}. As a consequence, the embedding potential becomes inaccurate when the subsystems feature a large overlap between their electron densities \cite{genova2014,fux2008,kiew2008} (this is because the larger the density overlap is, the larger the magnitude of the nonadditive potentials become). In FDE, the subsystem KS equations are left to converge to selfconsistency with respect to each other. This is often achieved by employing the so--called freeze-and-thaw procedure \cite{weso1996b,jaco2008b} (as done in ADF and other molecular codes) or via updating the embedding potential at every SCF cycle as done in CP2K \cite{iann2006,lube2014} and Quantum-Espresso \cite{genova2014,fdeinqe,krish2015a}. It is worth noting that FDE scales linearly with the number of subsystems provided that linear scaling methods for the solution of the electrostatic problem are employed \cite{jaco2008b}.

\begin{figure}
\includegraphics[width=0.8\textwidth]{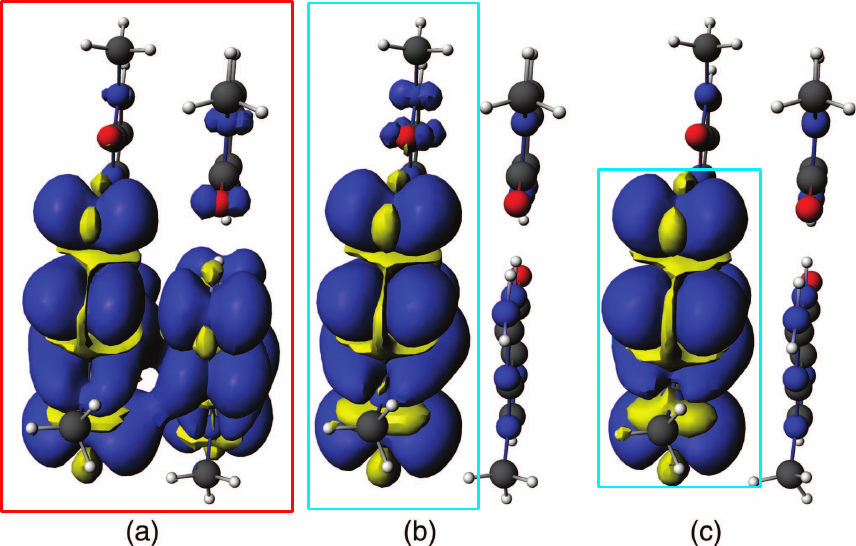}
\caption{\label{fde1}Spin densities of a guanine-cytosine dimer radical cation, $(GC)_2^+$. (a) KS-DFT supramolecular calculation using PW91 functional, (b) FDE calculation considering two subsystems where the left side subsystems (blue contour) is positively charged and (c) FDE calculation for four subsystems with one subsystems (blue contour) is positively charged. The nucleobases structures and spin densities were taken from ref \citenum{pava2011b}.}
\end{figure}

The earliest example of diabatization by FDE was given in Ref.\cite{pava2011b}. This is shown in Figure \ref{fde1}, where the spin densities for a pair of guanines are calculated. KS-DFT of the supersystem carried out with semilocal XC functionals fails in the prediction of the spin density. This is because the self-interaction error makes the spin density spread on both guanines against the prediction given by more accurate theoretical work\cite{voityuk2006a} and experimental studies\cite{kumar2011,mantz2007,burin2008}. On the contrary, FDE localizes the charge on a guanine of choice.  

The fact that FDE was able to provide  subsystem-localized electronic structures was known since its early application to systems with unpaired electrons \cite{neug2005f,weso1999,kevo2013,solo2012}. Later, this ability of FDE was explored for the computation of diabatic states for electron transfer \cite{pava2011b,pava2013a,ramo2014,ramo2015a} and to compute hyperfine coupling constants \cite{neug2005f,kevo2013}. 

The question that one can raise is why FDE calculations yield charge localized states? We provide here four reasons \cite{pava2013a,ramo2015a}. 
\begin{enumerate}
 \item Orthogonality is not imposed between the molecular orbitals belonging to different subsystems. 
 \item FDE calculations can be executed in the monomer basis set. This is known as FDE(m) method \cite{jaco2007}.
 \item FDE calculations are always initiated with a subsystem localized initial guess of the electron density.
 \item Electrons of a subsystem, remain localized also because there are repulsive walls in the region of the surrounding (frozen) fragments.
\end{enumerate}

The first reason, is important because it directly removes a bias towards delocalization which results due to orthonormalization of the molecular orbitals, as already noted by Dulak and Wesolowski \cite{dula2006}. 
The second and third reasons come together, the lack of  basis functions on the surrounding subsystems, does not allow substantial charge transfer between the subsystems. As a consequence, the SCF is biased to converge to localized electronic structures.

\begin{figure}
\includegraphics[width=0.8\textheight]{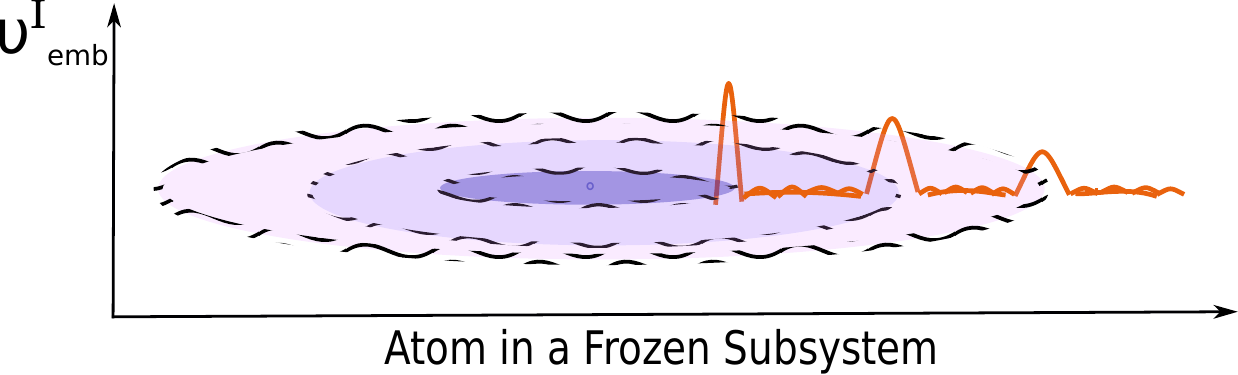}
\caption{\label{fde2} Exemplification of the embedding potential at the atomic shells of the surrounding subsystems. Figure from reference \citenum{ramo2015a}. }
\end{figure}

The fourth reason makes reference to the approximate nature of the term $\frac{\delta T_{\rm s}[\rho]}{\delta\rhor}-\frac{\delta T_{\rm s}[\rhoi]}{\delta\rhoir}$ (also known as nonadditive kinetic energy potential which is part of the embedding potential) in the region of the frozen fragments (e.g.\ in the region where $\rho_J$ with $J\neq I$ is larger than any other subsystem electron density). Approximate nonadditive kinetic energy potentials fail in canceling out the attractive potential due to the nuclear charge in the vicinity of the nucleus of the surrounding frozen subsystems \cite{jaco2007,wang2000}, and they do not reproduce the exact potential at intermediate regions \cite{thak1992,perd1992,fux2010}, especially in the vicinity of an atomic shell \cite{wang2000,fux2010}. In that region they cross the exact potential and large potential walls arise. A simplified depiction of this effect is devised in figure \ref{fde2}. 

In this scenario, diabatic states can be generated with FDE by performing at least two simulations, one featuring a hole/electron on the donor while the acceptor is neutral and one calculation in which the charge hole/electron is on the acceptor. The result is two charge localized states, whose, densities and Kohn-Sham orbitals are used in a later step in order to build the diabatic Hamiltonian and overlap matrices, needed to compute the diabatic coupling matrix element.

\subsubsection{FDE-ET method}
\label{fdeetm}
\begin{figure}
\includegraphics[width=0.8\textheight]{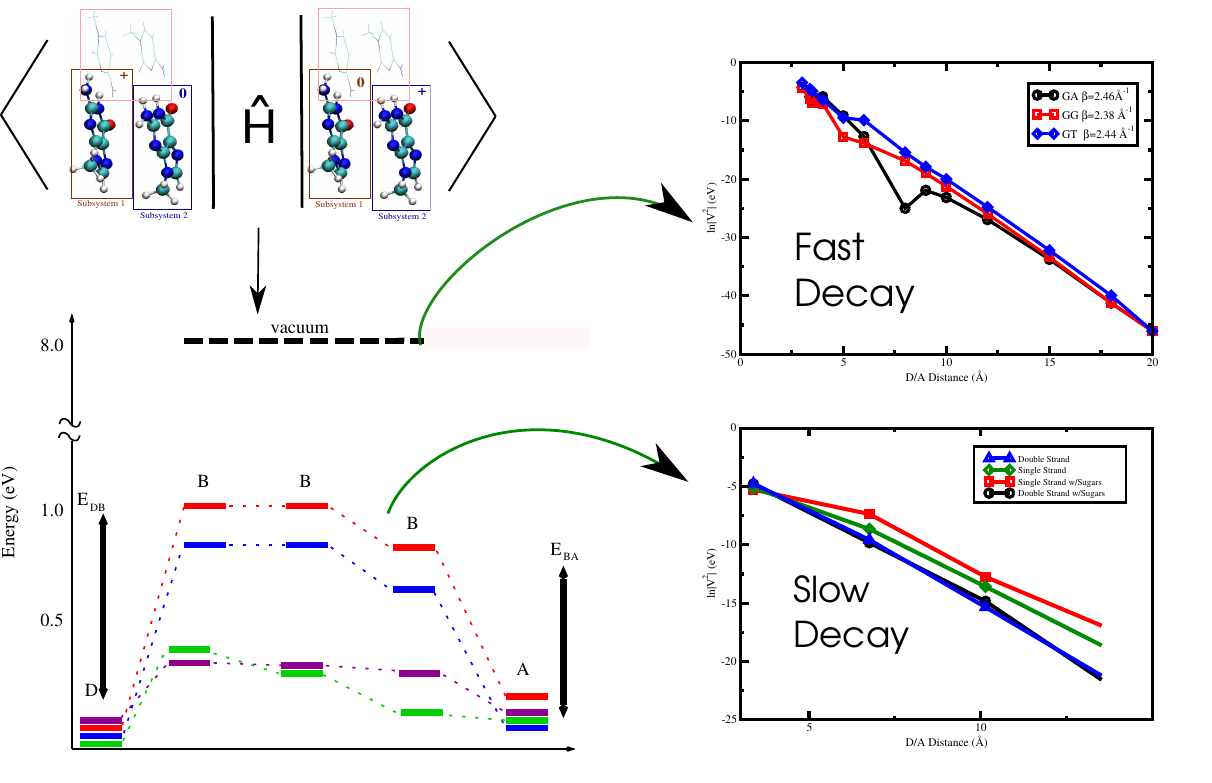}
\caption{\label{dia1} Energy dependence of a charge transfer process. The off-diagonal element (Dirac notation) will decay as the potential well that the charge has to overcome increases. Two cases: for vacuum as a potential well we have a faster decay, and when molecules act as a bridge the transfer will decay slower.}
\end{figure}

FDE-ET is a methodology which computes Hamiltonian couplings from diabatic states generated by an FDE calculation. Electron transfer reaction are usually described in the basis of a two-state formalism\cite{newton1991}, taking as basis set two broken-symmetry charge-localized states. This methodology can also approach models for the superexchange mechanism\cite{nitzan_book,Lewis01081997,bixo1999,jort1998,rena2013,gies2000}, where the transfer is still modelled by a Two-dimensional basis set but the coupling includes the effect of non-resonant bridges states. Figure \ref{dia1}, illustrates the difference between tunneling through the vacuum and through a set of bridge states. The bridge could be comprised of one or more molecules, a covalent bond or any other type o potential barrier as long as its height is lower than the one when vacuum separates donor and acceptor. As it is shown in Figure \ref{dia1}, the higher the potential barrier the faster the coupling decays with respect to the donor--acceptor distance.

In FDE-ET we seek a method capable of computing the Hamiltonian matrix in the basis of charge--localized states generated with FDE. First, we have to define the needed matrix elements.
As diabatic states are not the eigenfunctions of the molecular Hamiltonian, the off-diagonal elements of such Hamiltonian are not zero and can be approximated by the following formula \cite{thom2009a,fara1990} if $\psi_{D}$ \ and $\psi_{A}$ are slater determinants representing the donor and acceptor diabats:
\begin{equation}
\label{cou}
 H_{DA}=\langle \psi_{D} | \hat H |\psi_{A}\rangle=S_{DA}E\left[\rho^{(DA)}(\br)\right].
\end{equation}
Here $\hat H$ is the molecular electronic Hamiltonian, and $\rho^{(DA)}(\br)$ is the transition density defined as $\rho^{(DA)}(\br)=\langle \psi_D | \sum_{k=1}^{n_e} \delta (\br_k-\br) | \psi_A\rangle$, with $n_e$ being the total number of electrons in the system (i.e.\ the sum of the electron number of all subsystems) and $E\left[\rho^{(DA)}(\br)\right]$ is an energy density functional. The donor--acceptor overlap matrix elements are found by computing the following determinant:
\begin{equation}
\label{eq:s}
 S_{DA}=\mathrm{det}\left[\mathbf{S}^{(DA)}\right],
\end{equation}
where $\bS^{DA}_{kl}=\langle \phi_{k}^{(D)} | \phi_{l}^{(A)} \rangle$ is the transition overlap matrix in terms of the occupied orbitals ($\phi_{k/l}^{(D/A)}$)\cite{maye2002,thom2009a}. Thus, the transition density is now written in the basis of all occupied orbitals which make up the diabatic states  $\psi_{D}$ \ and $\psi_{A}$.
\begin{equation}
\label{Ivth}
 \rho^{(DA)} (\br)= \sum_{kl}^{\rm occ} \phi_{k}^{(D)} (\br)\left(\mathbf{S}^{(DA)}\right)_{kl}^{-1} \phi_{l}^{(A)}(\br).
\end{equation}
The Hamiltonian coupling is not $H_{DA}$, but it is generally reported as the coupling between the L\"{o}wdin orthogonalized $\psi_D$ and $\psi_A$. For only two states this takes the form,
\begin{equation}
\label{coulow}
 V_{DA} = \frac {1}{1-S_{DA}^2} \left(H_{DA}-S_{DA}\frac{H_{DD}+H_{AA}}{2}\right).
\end{equation}

Turning to the superexchange picture, the effective coupling, is a summation of the contribution given by the interaction between D and A and the interaction of D and A with all bridges states, namely:
\begin{equation}
\label{efcou}
 V_{DA}(E) = {\tilde V}_{DA} + \underbrace{\mathbf{\tilde V}_{DB}^{T} \mathbf{G}_{B}(E)\mathbf{\tilde V}_{BA}}_{V_{\rm bridge}},
\end{equation}
where the superscript $T$ stands for transpose, $\mathbf{G}_{B}(E)$ is the Green's operator, defined as
\begin{equation}
\label{green}
 \mathbf{G}_{B}(E) = -( \mathbf{\tilde V}_{B}-E~\mathbf{\tilde I}_B)^{-1},
\end{equation}
As shown in equation \ref{efcou}, $\tilde V_{DA}$ is the coupling for the donor-acceptor transfer, which in the absence of bridge states (CT through vacuum), would be the only contribution to $V_{DA}(E)$. On the other hand, if bridge states are present, the contribution to $V_{DA}(E)$ is given by the second addend in equation \ref{efcou}. Generally, $E$ appearing above is the energy at which the tunneling event occurs (i.e.\ at the crossing seam of the Marcus parabolas). In our works \cite{ramo2014}, $E$ was chosen to be in between $E_{A}$ and $E_B$, and specifically to be $\frac{E_D+E_A}{2}$. This choice is invoked by several works in the literature \cite{hatc2008,newton1991,marc1987,voityuk2012} where it is well known that there is a mild dependence of the coupling with the tunneling energy\cite{marc1987}. However, this equation holds when there is no resonance between D, A and the bridges states\cite{even1992,nitzan_book,newton1991,lowd1963,Larsson1981,Priya1996}. If near-degeneracies appear then the transport regime transitions to resonant tunneling or hopping. 

\subsubsection{Distance dependence of the electronic coupling}
In this section, we discuss calculations of the coupling matrix element ($V_{DA}$) of hole transfer from a donor to an acceptor molecule through the vacuum. This means that the initial state of hole transfer is the donor molecule (D), and the final state the acceptor molecule (A), and no intermediate bridge states are considered. Any reliable method for computing couplings should be able to reproduce high level calculations of CT coupling in small molecular dimers. For this purpose, we initially chose 23 biologically relevant $\pi$-stacks \cite{ramo2014}, in order to analyze the distance dependence of the coupling, separations of 3-20 \AA \ were considered, as result a total of 276 coupling calculations were ran. Overall, our couplings show a good agreement with previous computations (e.g.\ we reproduced the decay factors, $\beta$, $\pi$-stacked dimers separated by vacuum). 

When a test set for hole transfer couplings featuring high accuracy couplings became available \cite{Kubas2014} we could compare systematically the FDE-ET couplings with the benchmark values \cite{ramo2015a}. Benchmark calculations were ran on a set of 15 of $\pi$-stacked dimers. This study was rigorous, and tested the effect of the basis set size, nonadditive kinetic energy functionals (NAKE) and exchange-correlation functionals (XC) on the value of the computed couplings. The most important finding that resulted from the benchmark work resided in the fact that GGA functionals coupled with a medium sized basis set and the PW91k NAKE functional allow the FDE-ET method  to yield reliable electronic couplings as tested against high-level correlated wavefunction (MRCI+Q, NEVPT2 and SCS-CC2) methods applied to the array of dimers. The PBE and PW91 functionals are found to be a good choice in each case considered with a MAX error lower than 50 meV and an overall MRUE of a little over 7\% in both cases\cite{ramo2015a}.
Statistically, we found that hole transfer couplings are relatively insensitive to the choice of NAKE functionals, while our analysis of the basis set dependence shows that QZ4P basis set is the most problematic, as it often biases the FDE convergence to nonphysical states at short intersubsystem separations -- a problem already well documented in the FDE literature\cite{Fradelos2011,Fradelos2011b}.
Finally, Table \ref{bests} compares the performance of FDE-ET for different levels of theory. The results for GGAs are in good agreement with the benchmark values, and in some cases they showed to be superior to hybrid and meta-GGA functionals, particularly PBE and PW91. B3LYP also stands out as another valuable choice.

Generally, all functionals perform well in the FDE-ET coupling calculations making FDE-ET a method that is relatively insensitive to the XC and NAKE functional choice.
\begin{table}[htp]
\begin{center}
\begin{tabular}{crrr}
 \toprule

Set & MUE(MeV) & MRUE(\%) & MAX(meV)\\
 \hline
PBE/PW91k/TZP    & 15.3 & 7.1 & 49.6\\
PW91/PW91k/TZP   & 15.2 & 7.1 & 49.1\\
B3LYP/PW91k/TZP  & 18.1 & 7.9 & 58.5\\
M06-2X/PW91k/TZP & 18.0 & 8.2 & 54.9\\
 \bottomrule
\end{tabular}
\end{center}
\caption{Mean statistical values for the best XC-functional choices. PW91k is the NAKE along this work. Reproduced with permission from reference \citenum{ramo2015a}}
\label{bests}
\end{table}

\subsubsection{Hole transfer in DNA oligomers}
\label{sect:DNA}
\begin{figure}[ht]
\begin{center}
\includegraphics[width=0.8\textwidth]{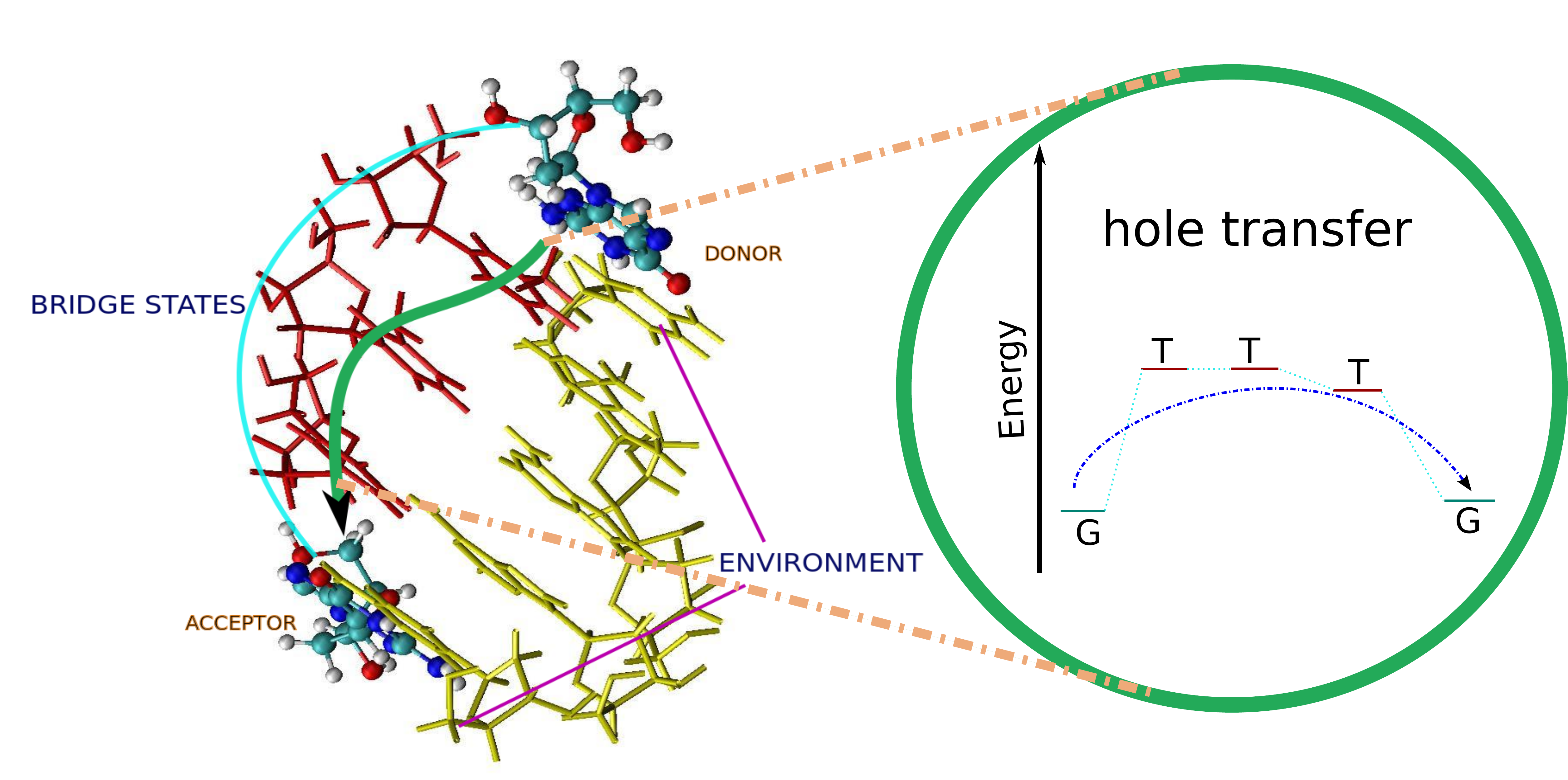}
\end{center}
\caption{The dephosphorilated \sgtg\ B-DNA oligomer employed in the hole transfer coupling calculations. As the figure depicts, the hole tunnels from the bottom guanine (in balls and sticks) to the top guanine. The tunneling wall is provided by a series of three thymines (red branch, labeled as ``bridge''). The counterstrand, C(A)$_N$C, acts as a solvating environment (in yellow, labeled as ``spectators'') and no hole is allowed to localize on it.}
\label{dna1}
\end{figure}

In this section, we discuss an interesting application of FDE-ET to charge transfer in biosystems.The electronic coupling for hole transfer in a completely dry B-DNA structure of \sgtg \ and \sgag \ was calculated. The structures considered lack water molecules, metal counterions and phosphate linker groups. The latter is because the applicability of FDE is restricted to non-covalently bound molecular fragments. Consequently, appropriate modifications to the B-DNA structure had to be made: we have removed the phosphate groups and capped the dangling bonds with hydrogen atoms at 1.09 \AA \ from the bonding atom. The resulting structure of the modified \sgtg \ is depicted in Figure \ref{dna1}. The largest system considered is the double strand with ribose groups and counts 308 atoms and 1322 electrons. In this study, the role of the environment on the CT in DNA is elucidated and analyzed on the basis of an all-electron computation. 

Regarding the energetics (site energies), an uneven stabilization of the bridge states compared to donor/acceptor states occurs in both type of oligomers, being this effect more pronounced in the \sgtg \ system than in the \sgag \ system. By inspection of the overall electrostatics of the interaction between G:C and T:A \cite{mill1990}, we notice that T has a strong permanent dipole pointing towards A, similarly to C:G. Instead, A has a much weaker dipole compared to C or T and thus upon contact of the GTG strand with the CAC strand the cytosines will stabilize much more the holes on Gs than the adenines can stabilize the holes on Ts, hence the tunneling wall increases from single strand to double strand.

Regarding the couplings, when the magnitude of the through space and through bridge couplings are inspected, our calculations show that the effects of the ribose groups and the nucleobases in the counterstrand are opposite and different in magnitude depending on the oligomer size (see table \ref{tab3}). We conclude, however, that the effect of the counterstrand on the computed superexchange couplings completely overpowers any effect due to the presence of the ribose groups.
\begin{table}[htp]
\begin{center}
\begin{tabular}{lrrrr}
\toprule
 & ${\tilde V}_{DA}$ (meV) & $V_{\rm bridge}$ (meV) & E$_{\rm DB}$ (eV) & E$_{\rm BA}$ (eV) \\
\hline
\multicolumn{5}{c}{{\sc Single Strand no Ribose}} \\
\hline
GG        & 78.13   &      &       &     \\
GTG       & 0.76   & 12.46  & 0.71   & 0.50 \\
G(T)$_2$G & 0.01   &  1.13  & 0.79   & 0.66 \\
G(T)$_3$G & --  &  0.09  & 0.79   & 0.77 \\
\hline
\multicolumn{5}{c}{{\sc Double Strand no Ribose}} \\
\hline
GG        & 92.6   &      &        &     \\
GTG       & 0.65 & 7.66   & 0.93   & 0.96 \\
G(T)$_2$G & 0.01 & 0.47   & 1.11   & 0.94 \\
G(T)$_3$G & -- & 0.02   & 0.99   & 1.16 \\
\hline
\multicolumn{5}{c}{{\sc Single Strand with Ribose}} \\
\hline
GG        & 71.38   &      &        &     \\
GTG       & 0.18 & 25.01  & 0.43   & 0.37 \\
G(T)$_2$G & 0.02 &  1.70  & 0.58   & 0.37 \\
G(T)$_3$G & -- &  0.21  & 0.41   & 0.41 \\
\hline
\multicolumn{5}{c}{{\sc Double Strand with Ribose}} \\
\hline
GG        & 91.07   &      &        &     \\
GTG       & 0.02 & 7.35   & 0.62   & 0.87 \\
G(T)$_2$G & 0.02 & 0.61   & 0.93   & 0.60 \\
G(T)$_3$G & -- & 0.02   & 0.50   & 0.82 \\
\bottomrule
\end{tabular}
\end{center}
\caption{Through-space and through-bridge electronic coplings and tunneling energy gaps for single and double strand \sgtg\ B-DNA, including the effects of the backbone (sugars). A -- is shown for values below 0.01meV. Reproduced with permission from reference \citenum{ramo2014}.}
\label{tab3}
\end{table}

\subsection{Constrained Density Functional Theory Applied to Electron Transfer Simulations}
Alternatively to FDE-ET methodology, constrained DFT (CDFT hereafter), a DFT--based procedure that was initially proposed by Dederichs et al \cite{dede1984}, and later introduced by Van Voorhis and Wu\cite{wu2005} with the aim of applying it to charge transfer reactions. CDFT is an effective method for calculating diabatic states for electron transfer, it relies on the idea of seeking the ground state of a system subject to a constraint. This can be achieved by adding to the conventional KS Langragian an additional term that accounts for the constraining external potential, this reads as\cite{kadu2012}:
\begin{equation}
\label{cdft1}
\mathcal{L}_{CDFT}[\rho] =\underbrace{\mmfunc{E}{HK}{\rho} + \int \pot{ext}{(\br)} \rhor \d\br - \mu \left[ \int \rhor \d\br - N_e  \right]}_{\mathrm{same~as~regular~KS-DFT}} + \     V_c\left[ \int \omega_c(\br) \rhor \d\br-N_c\right] 
\end{equation}
where $V_c$ is the Lagrange multiplier of the constraint, $\omega_c(\br)$ acts as the weight function that defines the constraint, typically a population analysis based on a real-space\cite{vanv2010a} partitioning (such as Becke pop. analysis). $N_c$ is the value of the constraint, and at self consistency it should satisfy the following tautology:
\begin{equation}
\label{cdft_const}
 N_c = \int \omega_c(\br) \rhor \d\br
\end{equation}
Having defined the constraint parameters, the energy of the system can be computed by solving the KS equation for the constrained system:
\begin{equation}
 \label{cdft2}
 \left( -\frac{1}{2}\nabla^2 + \int \frac{\rho(\br')}{|\br-\br'|}d\br' + \upsilon_{xc}(\br) + V_c \omega_c (\br)\right) \phi_i[V_c](\br) = \epsilon_i[V_c] \phi_i[V_c] (\br)
\end{equation}
where we have emphasized the functional dependence of the orbitals and orbital energies to the CDFT Lagrange multiplier. Clearly, the integral in \eqn{cdft_const} is only satisfied when an appropriate choice of $V_c$ is employed. The term $\upsilon_{xc}$ is the exchange--correlation potential and $\phi_i$ are the KS--orbitals. Note that $\rho (\br) \equiv 2 \sum_i |\phi_i(\br)|^2$ for closed shell systems. Thus also the density is a functional of the CDFT Lagrange multiplier. To our knowledge, the CDFT algorithm can be found on NWChem\cite{nwchem_program}, Q-Chem\cite{qchem}, CPMD \cite{ober2010}, PSI \cite{Evan2013}, SIESTA \cite{souz2013}, and ADF \cite{ramo2015b}.
Computing the electronic coupling on a diabatic basis can be carried out similarly to \eqs{cou}{coulow} or using a CDFT-specific prescription \cite{wu2006}. 

\begin{figure}[ht]
 \includegraphics[width=0.6\textwidth]{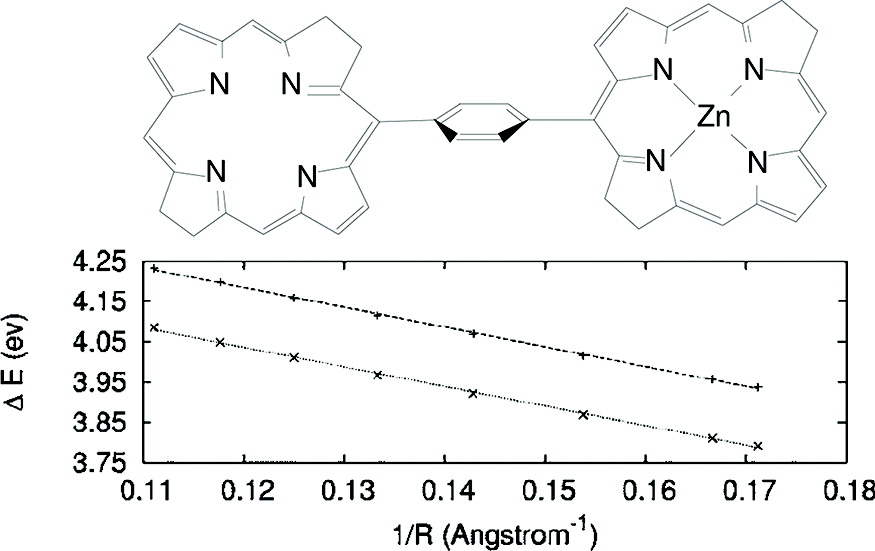}
  \caption{charge-transfer state energies of ZnBC-BC as compared to its ground-state energy at 5.84 \AA separation. Lower line, $Zn^+BC^-$;upper line, $Zn^-BC^+$. Taken from ref.\citenum{wu2005}}
 \label{cdft5}
\end{figure}

An example is the long range charge transfer excited states of the zincbacteriochlorin-bacteriochlorin complex (ZnBC-BC), an important structure in photosynthetic process in bacteria, has been calculated on the basis of CDFT procedure\cite{wu2005,wu2006}. In figure \ref{cdft5}, the excited states at different intersubsystem distances is depicted, where the last point of each curve represent the CT excitation energy of the linked complex. These energies are in good agreement with previous methodologies\cite{dreu2004}, and also demonstrates that by constraining CDFT ground state the excitations are more accurate than TDDFT energies (1.32-1.46 eV)\cite{wu2005}.

Additionally, CDFT can generate states with partial charges\cite{wu2005}, this is of particular importance, for example in metal--ligand CT processes, where the diabatic states can be generated by constraining the charge on the ligand and metal center. 

Recently, the CDFT implementation of CPMD was tested against high-level wave function methods in the computation of electronic couplings for hole and excess electron transfer \cite{Kubas2014,Kubas2014b}. CDFT was shown to be on average within 5.3\% of the benchmark calculations if 50\% HF exchange was introduced (the average deviation goes up to 38.7\% if HF exchange is not used).

\subsection{Fragment Orbital DFT}
The fragment orbital DFT or FODFT is a computationally low-cost method to calculate electronic couplings. This is because the wavefunctions of each diabatic state are approximated by the fronteer orbitals of the isolated donor/acceptor fragments \cite{senn2003,kuba2013,ober2012}. The underlying approximations in FODFT are that (1) the interactions between donor and acceptor have not effect on the orbital shape, (2) the coupling component related to orbitals below the fronteer is neglected (e.g.\ frozen core).
In FODFT, the wavefunctions can be described by a single determinant of $N-1$ spin-orbitals $\phi$, where $N=N_A+N_D$ i.e.\ the sum of the number of electrons of the neutral donor and acceptor. These determinants are built from the KS orbitals of the noninteracting isolated donor and acceptor fragments.
\begin{align}
 \label{fodft1}
  \nonumber
 &\psi_a \approx \psi_a^{D^+A} = \frac{1}{\sqrt{(N_D-1+N_A)}!} \mathrm{det}\left( \phi^1_D\dots \phi^{N_D-1}_D \phi^1_A\dots\phi^{N_A}_A\right)\\
 &\psi_b \approx \psi_b^{DA^+} = \frac{1}{\sqrt{(N_D+N_A-1)}!} \mathrm{det}\left( \phi^1_D\dots \phi^{N_D}_D \phi^1_A\dots\phi^{N_A-1}_A\right)
\end{align}
The Hamiltonian used to calculate the CT matrix elements is the KS-Hamiltonian. Namely,
\begin{align}
 \label{fodft2}
 \nonumber
 &H^{KS}_a = \sum_{i=1}^{N_D+N_A-1} h^{KS}_{a,i}\\
 &H^{KS}_b = \sum_{i=1}^{N_D+N_A-1} h^{KS}_{b,i}
\end{align}
where $h^{KS}_{a,i}$ are the one--particle KS--Hamiltonians for either the "a" diabat or the "b" diabat. One feature of these Hamiltonians is that they are state dependent, thus, they are made of the combination of orbitals of donor and acceptor species at the given state. The transfer integral, or coupling between states, is calculated as:
\begin{align}
  \label{fodft3}
 \nonumber
 H_{a,b} =& \langle \psi_a|H|\psi_b\rangle\\
 \nonumber
          \approx& \langle \psi_a^{D^+A}|H^{KS}_a|\psi_b^{DA^+}\rangle\\  
          \approx& \langle \phi^N_a|h^{KS}_{a,i}|\phi^N_b\rangle
\end{align}
Where $N$ above is the fronteer orbital for $D$ or $A$. Recently, Kubas et al.\cite{Kubas2014} have shown the differences of two FODFT flavors in the calculation of the hole transfer coupling for the HAB11 database. As we can see in Table \ref{fodft6}, the implementation including $N_A+N_D$ orbitals in the KS Hamiltonian (indicated by $2N$ in the table) as done in ADF \cite{senn2003} is more accurate than the implementation using one of the Hamiltonians in \eqn{fodft2} (which is indicated by $2N-1$ in the table).
\begin{table}[htp]
\begin{center}
\begin{tabular}{cccccc}
 \toprule
        &FODFT(2N-1)&	FODFT(2N)	&ADF(2N)	&FODFTB   &REF\\
\hline       
 Ethylene	&367.7	&389.2	&388.4	&343.7	&519.2\\
Acetylene	&316.9	&345.8	&345.3	&212.0	&460.7\\
Cyclopropene	&418.8	&443.7	&439.4	&367.4	&536.6\\
Cyclobutadiene	&323.3	&346.9	&345.6	&261.6	&462.7\\
Cyclopentadiene &343.3	&360.6	&358.7	&283.2	&465.8\\
Furane	        &315.6	&334.0	&333.7	&280.3	&440.3\\
Pyrrole	        &328.7	&347.8	&347.7	&286.2	&456.3\\
Thiophene	&341.2	&357.8	&356.1	&264.8	&449.0\\
Imidazole	&310.7	&328.9	&328.2	&277.5	&411.6\\
Benzene	        &342.4	&353.5	&354.1	&299.9	&435.2\\
Phenol	        &190.5	&211.3	&279.5	&231.4	&375.0\\
 \bottomrule
\end{tabular}
\end{center}
\caption{$H_{DA}$ (meV) calculated with various FODFT approaches for HAB11 dimers at intermolecular separation of 3.5 \AA. Reproduced with permission from table XI of Kubas et al.\cite{Kubas2014}.} 
\label{fodft6}
\end{table}

\begin{figure}[ht]
\begin{center}
\includegraphics[width=0.8\textwidth]{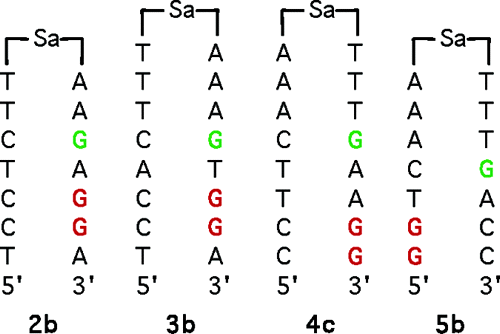}
\end{center}
\caption{DNA hairpins used to calculate hole transfer via superexchange mechanism between a single guanine (green) and a guanine doublet (red). Reproduced with permission from Senthilkumar et al.\ (Figure 1 of the reference) \cite{senth2005}.}
\label{fodft7}
\end{figure}

FODFT has been successfully applied to models of CT in molecular semiconductors \cite{coro2007,gajd2013} and also for modeling CT in biosystems. In the following, we provide applications of FODFT to biological CT: such as the determination of the hole rates on DNA hairpins linked by stilbenedicarboxamide (Figure \ref{fodft7}), and the elucidation of the electron transfer between two cofactors in the SO enzyme. 

\subsubsection{Hole transfer rates on DNA hairpins}

The absolute rates were determined by using Marcus theory, in \eqn{marcus}, where the electronic coupling was calculated according to FODFT and the superexchange regime (see section \ref{fdeetm}). Knowledge of the rates forward and backward (see table \ref{fodft8}) enables one to determine the equilibrium constant $K=k_t/k_{-t}$ and the free energy change $-\Delta G=-k_BTln(K)$. Comparable results with experimental results \cite{lewis2003} were obtained.

\begin{table}[htp]
\begin{center}
\begin{tabular}{ccc}
 \toprule
 Sequence & $K=k_t/k_{-t}$ & $-\Delta G(eV)$\\
 \hline
2b & 3.5 &0.032\\
   &(7.5) &(0.052)\\
3b &39.8 &0.093\\
   &(6.7) &(0.049)\\
4b &20.0 &0.076\\
   &($>0.5$) &(-0.02)\\
5b  &10.0 &0.058\\
   &($>3$) &($>0.028$)\\
 \bottomrule
\end{tabular}
\end{center}
\caption{Equilibrium Constant, $K$, and Free Energy Change, $-\Delta G$, for hole transport between the proximal G site and the Distal GG Doublet. Experimental data in parentheses were taken from ref \citenum{lewis2003}. Table reproduced wth permission from ref \citenum{senth2005}} 
\label{fodft8}
\end{table}

\subsubsection{The Curious Case of Sulfite Oxidase}
An interesting and elusive candidate for electron transfer studies is the Sulfite Oxidase protein\cite{asta2012}. For this protein, theory predicts an electron transfer rate between the cofactors (a heme and a molybdenum complex) about two orders of magnitude lower than what is measured experimentally\cite{kawa2006}. To address this issue, Beratan et. al., using the Pathways model, suggested that the donor and the acceptor are joined together by a flexible tether\cite{utes2010}.  As the tether allows the two cofactors to come sufficiently close to each other, electron transfer occurs at the rate shown by experiment. A recent simulation of this mechanism was carried out so that the protein was taken out of equilibrium and positioned in a new folded state featuring a much decreased cofactor distance (about 10 \AA). However, recent pulsed electron paramagnetic resonance measurements\cite{asta2012} indicated that the distance between the cofactors is unchanged on average from the one available in the crystal structure (32 \AA).  To approach this problem using FODFT, the crystal structure of the protein is obtained and only the chains of the protein between the two cofactors are considered. The chains of interest are broken into individual molecules and treated as separate bridges.
\\
\begin{figure}[ht]
\includegraphics[width=0.8\textwidth]{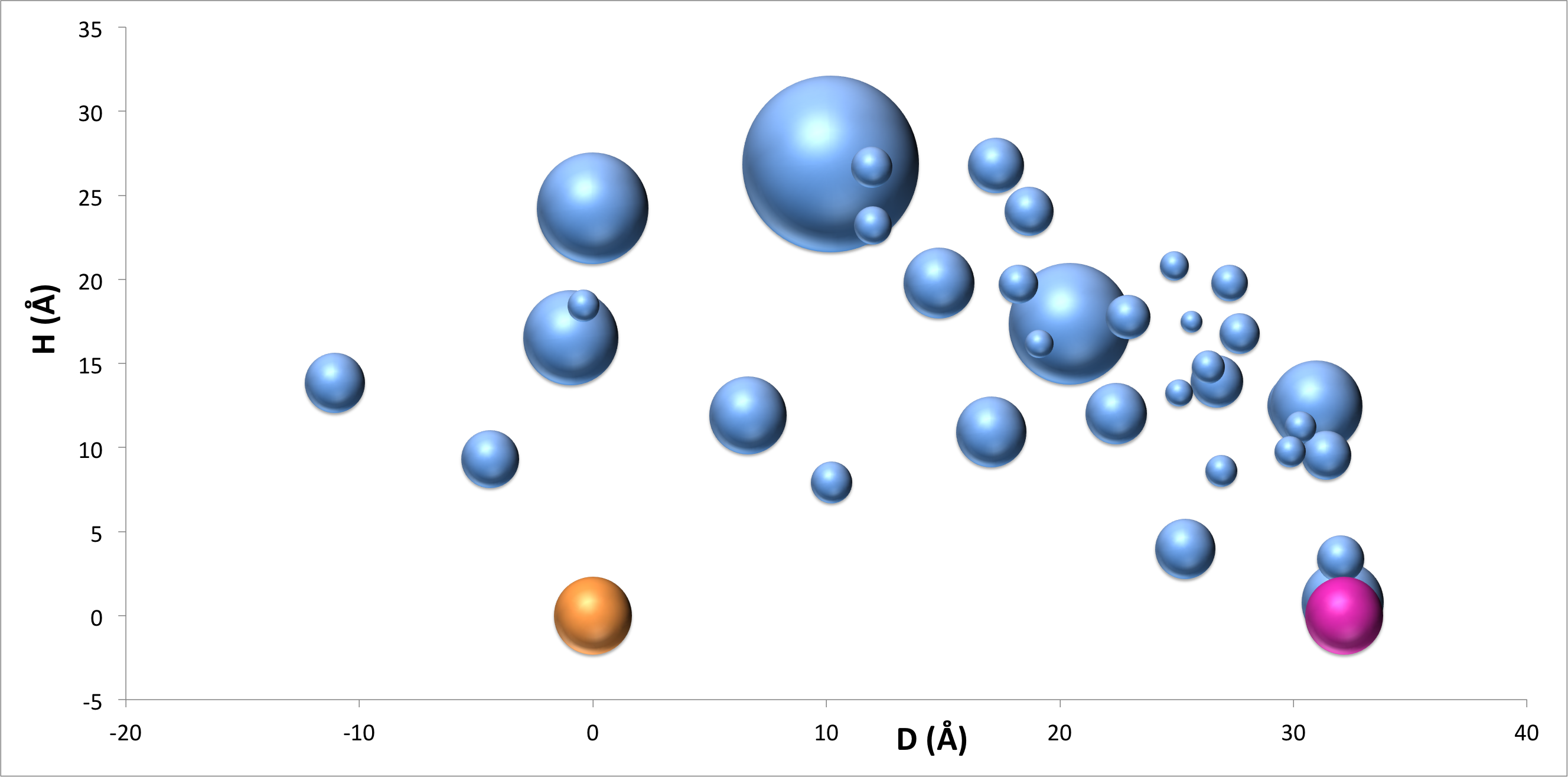}
\caption{Energy landscape for the hole transfer in Sulfite Oxidase: the circles represents the position of the center of mass of each fragment with respect to the electron transfer vector coordinate(the distance between the center of mass of the HEME complex (orange sphere) and the MOCO complex(pink sphere)); the size of the sphere is $1/(x^2)$ in which x is the difference between the HOMO energy of the fragment and the Fe (that is HOMO of Fragment - HOMO of Fe).  Fe and Mo were give a size of 1 for scale.}
 \label{so}
\end{figure}
\\
The FODFT computations that we present here will be part of a more in--depth study in a future publication\cite{ramo2015c}. Two ingredients are available from the simulations, the site energies, and the couplings between the sites. The energies of the hole transfer pathway for the electron transfer between the iron and the molybdenum is presented in Figure \ref{so}. With the aid of Koopman's theorem, the HOMO energies computed with FODFT are taken here as a measure of the ionization potential of each site. The simulation was able to shed light on some very interesting aspects of the couplings and the energy landscapes. Looking at the energy landscape in Figure \ref{so}, it is clear that a path with highest couplings can be directly drawn between the donor and the acceptor. The landscape also shows the possibility of hopping stations--molecules that exist between the donor and acceptor and are close to them in energy. These two aspects of the landscape alone hint the possibility of a hole transfer occurring over the 32 \AA. However, proteins are very complex structures with many variables such as size, dynamics and environment. Therefore, providing a quantitative analysis of the kinetic constant would require incorporating unbiased molecular dynamics and a more comprehensive structure to further characterize the role of these hopping stations.

\subsection{Ultrafast computations of the electronic couplings: The AOM method}
Recently, an ultrafast method to calculate electronic couplings was developed by Blumberger and coworkers\cite{Kubas2014b}. The analytic overlap method or AOM is a useful method if CT simulations need to be coupled with molecular dynamics, like in proteins\cite{beratan2008} or in organic semiconductors\cite{troi2006}. This quest requires hundreds or maybe thousands of $H_{DA}$ and site energy calculations. AOM offers an interesting alternative for such simulations. As in FODFT, AOM assumes that CT is only mediated by two SOMO orbitals (fronteer orbitals, similarly to FODFT), which correspond to each fragment. Then, small Slater type orbital basis for the valence states is generated. Thus the overlap integral is evaluated as follows:
\begin{align}
\label{aom1}
 \nonumber
 &S_{DA}=\langle \Psi_D|\Psi_A\rangle=\langle \phi_D^N |\phi_A^N\rangle\approx\\
 \nonumber
  &\approx\bar{S}_{DA}=\sum_{i\in D}^{atoms}\sum_{j\in A}^{atoms} c^*_{p\pi,i}c_{p\pi,j}\langle p_{\pi,i}|p_{\pi,j}\rangle
\end{align}
AOM further assumes contributions only from $p$-orbitals, particularly in organic compounds with $\pi-$conjugation, the p--orbital considered is that one perpendicular to the plane of $\pi-$conjugation. 

\begin{figure}[ht]
 \includegraphics[width=0.8\textwidth]{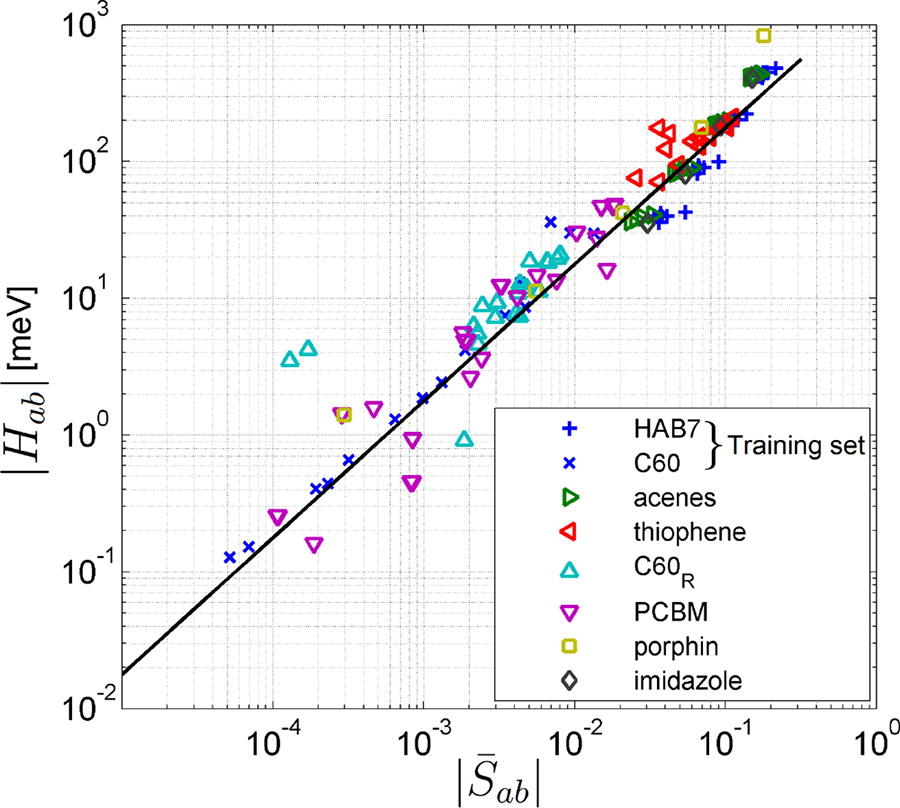}
 \caption{Correlation between electronic coupling matrix element from sFODFT and overlap between SOMO orbitals of donor--acceptor fragments. Taken from ref. \citenum{Kubas2014b}.}
 \label{aom2}
\end{figure}

Correlation of the overlap $\bar{S}_{DA}$ and the electronic coupling given by FODFT is shown in Figure \ref{aom2} for a set of dimers and their geometries. In this picture, a satisfactory linearity between these two parameters is witnessed, therefore reliable proportionality constants can be achieved. $H_{DA}=\bar{C}\bar{S}_{DA}$ is used in order to obtain the constant C that can be used to get couplings of similar compounds. This first version of this method shows transferability for homo-dimers. However, transferability when non equivalent donor and acceptor systems are considered needs to be explored. Nevertheless, AOM's speed makes it a very valuable option.
\subsection{Note on orthogonality}
\begin{figure}[ht]
 \includegraphics[width=0.8\textwidth]{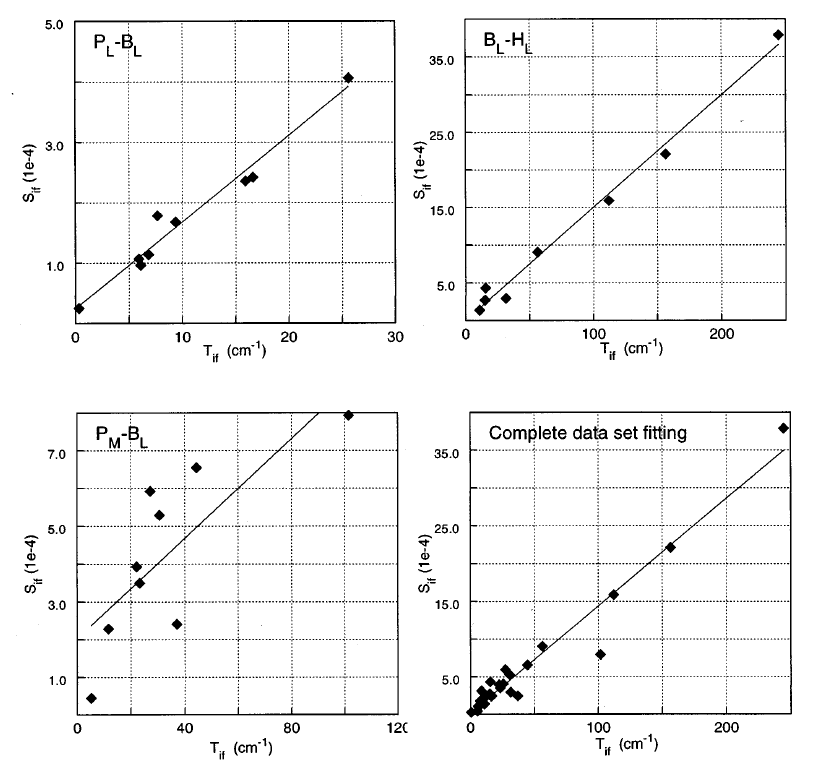}
 \caption{Correlation between electronic coupling matrix element and overlap between diabatic states for electron transfer. Taken from ref. \citenum{yu1997}.}
 \label{frie}
\end{figure}

When carrying out a large number of coupling calculations, one encounters all those low probability situations in which a method fails. In the case of FDE-ET, we probed a large number of so-called ``difficult cases''. Specifically, we faced two limitations of the FDE-ET method. If the diabats are orthogonal, or quasi orthogonal, numerical inaccuracies arise in the inversion of the transition overlap matrix in equation \ref{Ivth}. This is not specific to FDE-ET, but is a problem shared by all those methods that assume the diabatic states to be nonorthogonal \cite{newton1991,yu1997,cave1987,king1967,fara1990}. When they are orthogonal, some of the equations previously developed simply do not hold anymore. Yu et al.\ have applied equations similar to \eqs{cou}{coulow} and obtained a picture of the behavior of the electronic couplings in the photosynthetic reaction center, see Figure \ref{frie}.
If we concentrate on the left-side panels, we notice that in some cases the electronic coupling is proportional to the coupling, but in other cases (see lower left panel) the coupling seems to behave somewhat erratically as a function of the diabatic overlap.  To understand this, let us consider two distinct limiting cases: (1) orthogonality by symmetry considerations, and (2) spatial separation of the orbitals. We found that the second case is the predominant, as the distance between donor--acceptor increases the diabatic overlap becomes increasingly small. In the asymptotic limit \cite{migliore2011a}, there is a linear relationship between the coupling, the diabatic energy difference, and the diabatic overlap. If the overlap is small due to case (2), the asymptotic formula is not expected to hold. This explains the apparently contraddictory results presented in Figure \ref{frie}.

Regarding FDE-ET, both cases can be circumvented computationally by performing a singular value decomposition of the overlap matrix and then invert only those values which are larger than a threshold (i.e.\ Penrose inversion). For DNA presented in section \ref{sect:DNA}, the default inversion threshold of $10^{-3}$ was appropriate in most cases\cite{ramo2014}. However, three systems stood out: AG, GA and TT nucleobase pairs. All the systems above showed erratic behavior of the computed couplings for some specific donor-acceptor distances, specifically 4.0 \AA \ for AG, 3.5 \AA \ and 8.0 \AA \ for GA and 9.0 \AA \ for TT. We found that at those distances, the near singularity of the overlap matrix due to symmetry considerations (case 1 above) was the source of the erratic behavior. To circumvent these numerical issues, a threshold of $10^{-2}$ was adopted in these cases.

We thus conclude that although there is a formal relationship linking the diabatic overlap with the value of the coupling at large donor--acceptor distances \cite{migliore2011a}, generally assuming linearity in the coupling vs. overlap (as mentioned briefly in the previous section regarding the AOM method) can lead to large errors in the magnitude of the computed couplings. In the future, inspired by a recent work by Evangelista et al.\cite{Evan2013}, a more stable algorithm that invokes an orthogonalization first, and then the computation of the couplings will be developed in our group.

\subsection{A fully semiempirical method: Pathways}
Pathways \cite{Beratan31051991,kawa2006} is a semiempirical model which it is designed to reproduce electron transfer rates between cofactors in proteins.\cite{Beratan31051991,bera2002}. In essence, Pathways includes the contributions to the electronic tunneling from a stepwise path covering all nonbonded interactions, as well as the bonded ones at the nearest neighbor level. Namely: 
\begin{equation}
 \label{path1}
 |H_{DA}|^2= A^2\left(\prod_i \epsilon_i\right)^2
\end{equation}
where $\epsilon_i$ \ are the steps the charge need to make from donor to acceptor. For example, a hydrogen bond is one of such steps. The above product is maximized by searching all possible steps that contribute to the tunneling. The coupling is further split into three kinds of interactions:
\begin{equation}
 \label{path2}
 |H_{DA}|^2= A^2\left(\prod_i \epsilon_{bond}(i)\right)^2 \left(\prod_j \epsilon_{space}(j)\right)^2 \left(\prod_k \epsilon_{H-bond}(k)\right)^2
\end{equation}
Pathways can yield reliable predictions of the electronic couplings, where the CT process in proteins are mediated by the interactions of a single or multiple configurations that the protein can adopt\cite{Beratan31051991}.
Pathways has been successfully applied to a number of CT processes in protein environment. For instance, the electron transfer between the proteins cytochrome c2 (cytc2) and the photosynthetic reaction center (RC)\cite{Aquino1995277} in order to determine the protein structural dependence of this CT reaction, also, to look at the impact of structural and conformational variations on the electronic coupling between the proteins methylamine dehydrogenase and amicyanin from Paracoccus denitrificans\cite{Lande29062010}. 

\section{High-accuracy electronic couplings}

This section is devoted to describing those methods which are able to predict the electronic couplings accurately given a certain definition of the corresponding diabatic states. These methods start with a mathematical definition of diabatic states (usually a definition that involves localization of the electronic structure) such that the resulting states resemble the donor and acceptor states in the electron transfer reaction. Once this is achieved, an adiabatic-to-diabatic transformation matrix is generated which can be applied to the adiabatic Hamiltonian to result in the diabatic Hamiltonian featuring the sought electronic couplings in the off-diagonal elements. Usually, an accurate, wave function based level of theory is used for computing the adiabatic states and Hamiltonian \cite{cave1997}.
Examples of such techniques are, the Generalized Mulliken--Hush method developed by Newton and Cave\cite{newton1991,cave1997,subo2010}, Boys and Edmiston-Ruedenberg localizations of Subotnik et al.\cite{subo2010,vura2010,fate2013}, and fragment charge difference proposed by Voityuk and R\"{o}sch \cite{voityuk2002,hsu2009}. Their utility lies on the possibility of a very accurate computation of the corresponding adiabatic states, as was done for the hole transfer on $\pi$--stack DNA nucleobases at a CASPT2 and CASSCF level of theory accomplished by Voityuk et. al\cite{voityuk2006a}. That computation has served as the benchmark reference for many recently developed methodologies\cite{kuba2013,kumar2011,pava2013a,migliore2009b}.

Taking this as a motivation, let us briefly introduce each of the above methods, followed by some examples in which these methods where employed.

\subsection{GMH method}
In the two-state model, the charge localized diabatic states are related with the adiabatic states through the formula:
\begin{equation}
 \label{gmh1}
 E_{2,1}=\frac{1}{2}\left( E_D + E_A \pm \sqrt{(E_D-E_A)^2 + 4|H_{DA}|^2}\right)
\end{equation}
where $ E_{D/A}$ are the energies of the donor and acceptor states respectively, $E_1$ and $E_2$ are the energies of the adiabatic states, this means the energy of the ground state ($E_1$) and the first excited state ($E_2$). $E_1$ and $E_2$ can be obtained with any quantum chemistry method, however when highly accurate wavefunctions methods are employed also the resulting couplings will be of high quality. We now distinguish two cases: a symmetric case, for instance homo--dimers, and the general asymmetric case. In the symmetric case, we have that $E_D = E_A$, and thus the electronic coupling does not depend on the diabatization procedure. Namely,
\begin{equation}
 \label{gmh2}
 2|H_{DA}|=\Delta E_{12}
 \end{equation}
$\Delta E_{12}$ \ is the difference on energy between the adiabatic ground state $E_1$ \ and the first adiabatic excited state $E_2$. 

For asymmetric cases the GMH method prescribes that the proper diabatic states are those that diagonalize the adiabatic dipole moment matrix. In the two--state problem this is calculated as follows:
\begin{equation}
 \label{gmh3}
 |H_{DA}|=\frac{|\mu_{12}|\Delta E_{12}}{\sqrt{(\mu_{11}-\mu_{22})^2 +4\mu_{12}^2}},
\end{equation}
where $\mu_{ij}=\langle \Psi_i | \mu_{ET} | \Psi_j \rangle$, with $\mu_{ET}$ being the dipole moment in the direction of the electron transfer.
The power of GMH lies on the way one calculates the adibatic states. As we have seen through all sections is that the authors benchmark their own method by calculating proper diabatic couplings by using ab initio methods as multireference CI\cite{Kubas2014} and CASPT2 \cite{voityuk2006a}. 

As an example of this method, let us discuss the very first example given by Cave and Newton on their paper\cite{Cave1996}. The system $Zn_2H_2O^+$, the transfer of a hole is done over the Zn atoms. However, the water molecule, which is located at a fixed distance opposite to the Zn distance, causes an energy splitting of the Zn orbitals. Thus the electronic coupling is determined for the following diabatic states:
\begin{align}
 \label{gmh4}
 &Zn(^1S) + Zn^+(^2S)\rightarrow Zn^+(^2S) + Zn(^1S) \ s\rightarrow s'\\
 &Zn(^1S) + Zn^+(^2S)\rightarrow Zn^+(^2S) + Zn(^1S) \ p\rightarrow p'\\
 &Zn(^1S) + Zn^+(^2S)\rightarrow Zn^+(^2S) + Zn(^1S) \ s\rightarrow p' \\
 &Zn(^1S) + Zn^+(^2S)\rightarrow Zn^+(^2S) + Zn(^1S) \ p\rightarrow s'
\end{align}
where the transition are between the orbitals of the diabatic states (prime correspond to the acceptor state) (s-s'), (p-p'), (s-p') and (p-s'). CASSCF wavefunction method was used in all calculations. In Table \ref{gmh5} we collect the values for the different couplings and $\beta s$. There the analysis of the distance dependence of the coupling is carried out for several $R_{OZn}$ distances. Note that for an infinite $R_{OZn}$ distance, the couplings for (s-p') and (p-s') are equal. 

\begin{table}
 \begin{center}
  \begin{tabular}{ccccc}
   \toprule
  $r_{ZnZn}$(\AA)& $H_{ss'}$ & $H_{pp'}$ & $H_{sp'}$ & $H_{ps'}$ \\
   \hline
   && (a)&&\\
   4.0 &28.3 &23.6 &50.4 &42.7\\
   5.0 &10.5 &13.0 &51.7 &22.3\\
   6.0  &3.73 &7.55 &41.1 &10.1\\
   7.0  &1.09 &4.23 &21.8 &4.08\\
   8.0 &0.340 &2.57 &13.8 &1.62\\
   9.0 &0.0958 &1.44 &7.36 &0.611\\
   $\beta$& 2.28 &1.11 &0.81 &1.71\\
   && (b)&&\\
 4.0 & 29.7 &34.4 &59.3 &41.7\\
 5.0 & 7.95 &14.7 &38.5 &22.1\\
 6.0 & 2.34 &7.83 &19.1 &9.44\\
 7.0 & 0.698 &4.25 &9.56 &3.84\\
 8.0 &  0.203 &2.27 &4.78 &1.51\\
 9.0 & 0.0558 &1.16 &2.32 &0.574\\
 $\beta$&  2.49 &1.32 &1.32 &1.74\\
\bottomrule
  \end{tabular}
 \end{center}
 \caption{(a) Electronic coupling elements vs distance ($r_{ZnZn}$) for $Zn_2H_2O^+$ with $r_{ZnO}=2.05$ \AA. (b) Electronic coupling elements vs distance
($r_{ZnZn}$) for $Zn_2H_2O^+$ with $r_{ZnO}=3.05$ \AA. Results are in milihartree, the $\beta$ \ values were calculated on the range of 5-9 \AA. Taken from reference \citenum{cave1997}.}
 \label{gmh5}
\end{table}
We refer the reader to other publications which have evaluated the GMH method in detail in regards to its suitability in modeling two-state as well as multi-state problems for both excitation energy transfer and electron transfer processes \cite{yang2013,hsu2009,subo2008}.

\subsection{Other adiabatic-to-diabatic transformation methods}

Inspired by GMH, the electronic coupling can generally be obtained by rotating the corresponding adiabtic states into a set of diabatic states. Thus, each diabatic state can be expressed as a linear combination of rotated adiabatic states as\cite{subo2010}:
\begin{equation}
 \label{ld1}
 |\Xi\rangle=\sum_{j=1}^{N_{states}}|\Phi_j\rangle U_{ji}
\end{equation}
Under specific assumptions of the nature of the system--bath interaction (the following is valid for the condensed phase), the coupling can be estimated by constructing diabatic states based on Boys, Edminton--Ruedenberg (ER) or von Niessen--Edminton--Ruedenberg (VNER) localizations.
In Boys diabatization, the bath exerts a linear electrostatic potential on the system, thus the rotation matrix can be found by minimizing the following localizing function\cite{subo2008,subo2009}:
\begin{equation}
 \label{ld2}
f_{Boys}(U)=f_{Boys}(\Xi)=\sum_{i,j=1}^{N_{states}}=|\langle\Xi_i|\mu_{ET}|\Xi_i\rangle-\langle\Xi_j|\mu_{ET}|\Xi_j\rangle|^2.
\end{equation}
Boys localization was shown to be equivalent to GMH for CT reactions \cite{subo2008}.

ER diabatization, dictates that the bath exerts an electrostatic potential that responds linearly to the field generated by the molecular system (sum of donor and acceptor) system:
\begin{equation}
 \label{ld3}
 f_{ER}(U)=f_{ER}(\Xi)=\sum_{i=1}^{N_{states}}\int d\mathbf{r}_1 \int d\mathbf{r}_2 \frac{\langle\Xi_i|\hat{\rho}(\mathbf{r}_2)|\Xi_i\rangle-\langle\Xi_j|\hat{\rho}(\mathbf{r}_1)|\Xi_j\rangle}{|\mathbf{r}_1-\mathbf{r}_2|}.
\end{equation}

In VNER diabatization, the  bath exerts an electrostatic potential that responds linearly to the field of the total system, but the interaction potential is a Dirac delta function:
\begin{equation}
 \label{ld4}
 f_{VNER}(U)=f_{VNER}(\Xi)=-\sum_{i=1}^{N_{states}}\int d\mathbf{r}(\langle\Xi_i|\hat{\rho}^2(\mathbf{r})|\Xi_i\rangle-\langle\Xi_i|\hat{\rho}(\mathbf{r})|\Xi_i\rangle^2).
\end{equation} 

Just like GMH, once diabats are generated, the electronic coupling readily arises from the off-diagonal element of the electronic Hamiltonian and is equal to:
\begin{equation}
 \label{ld5}
 H_{DA}=\langle\Xi_D|H^{el}|\Xi_A\rangle
\end{equation}
 
\begin{figure}[ht]
\includegraphics[width=0.8\textwidth]{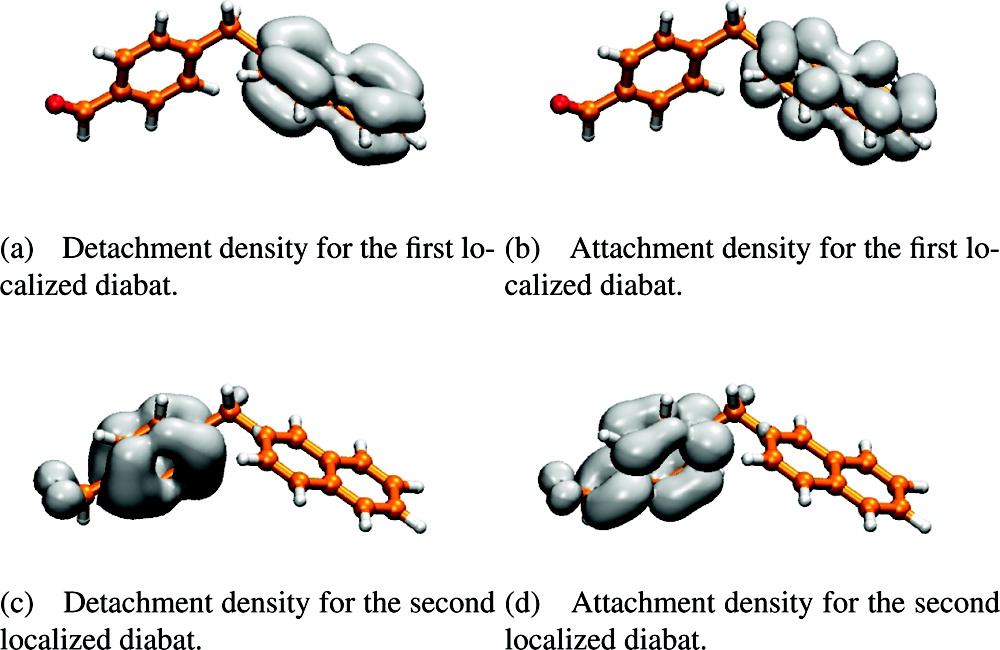}
\caption{Attachment/detachment plots for the occupied−virtual separated Boys localized diabatic excited triplet states near the avoided crossing. The molecule here is donor--$CH_2$--acceptor. Taken from ref. \citenum{subo2010}.}
 \label{ld6}
\end{figure}

For a series of bridge mediated excitation energy transfer experiments, where the donor is benzaldehyde and the acceptor naphthalene\cite{piotr1988,piotr1989}, the transfer rates and couplings were calculated using Boys diabatization. In Figure \ref{ld6} it is shown how well the orbitals are localized on either donor or acceptor edges in the various diabatic states. Although this computation does not concern a CT process, we want to stress the ability of this localization procedures in generating true diabatic states. 

\subsection{Fragment charge difference}
Similarly to the GMH method, the fragment charge difference (FCD) method yields a donor to acceptor coupling \cite{voityuk2002}:
\begin{equation}
 \label{fcd}
 |H_{DA}|=\frac{|\Delta q_{12}|\Delta E_{12}}{\sqrt{(\Delta q_{11}-\Delta q_{22})^2 +4\Delta q_{12}^2}}
\end{equation}
where $\Delta q_{1}$ and $\Delta q_{2}$ are the donor--acceptor charges differences in the respective adiabatic states $\psi_1$ and $\psi_2$. $\Delta q_{12}$ \ is the off diagonal term and is defined in a general form as $\Delta q_{ij}=q_{ij}(D)-q_{ij}(A)$, i.e.\ the difference of the populations of the transition charges.

Finally, when donor and acceptor are in resonance, i.e when $E_D=E_A$ \ or $\Delta q=0$, $H_{DA}=\frac{1}{2} \left( E_1-E_2\right)$. 

FCD method and its simplfied from (SFCD) are compared against GMH in the calculation of $H_{DA}$ for two Watson-Crick pairs $GC$ \ and $AT$. In Table \ref{fcd2}, the couplings calculated from FCD are in good agreement with GMH, the SCFD is also quite reasonable. Because the energy gap between donor and acceptor are large, the charge is completely localized on purines (lowest IP). However, if an electric field F (water molecule for instance) is tuned on near the pairs, the energy gap is then reduced and the coupling strength is enhanced. Overall, FCD is another good alternative to compute accurate couplings. However, the computation of the charges and the transition charges is dependent on the specific population analysis chosen. To our knowledge, only the Mulliken population analysis was used so far (i.e.\ the transition charges are evaluated on the basis of the MO coefficients over the atomic orbital basis set).
\begin{table}
 \begin{center}
\begin{tabular}{ccccccc}
 \toprule
    &           &           &            & GMH       & SFCD     & FCD \\
    &Basis set  & $E_2-E_1$ & $E_D-E_A$  & $H_{DA}$  & $H_{DA}$ & $H_{DA}$\\
 \hline
 GC  & 6-31G*    & 2.163     &2.159       & 0.0569    &0.0663    &0.0547\\
     &6-311++G** & 2.092     &2.086       & 0.0679    &0.0760    &0.0705\\
 AT  & 6-31G*    & 1.505     &1.502       & 0.0421    &0.0524    &0.0363\\
     &6-311++G** & 1.462     &1.459       & 0.0474    &0.0528    &0.0425\\
 \bottomrule
\end{tabular}
\end{center}
\caption{Hole coupling matrix element $H_{DA}$ of nucleobases within a Watson-Crick pairs. Energies in eV, dipole moment matrix elements in Debye, charges in a.u. Taken from ref. \citenum{voityuk2002}.} 
\label{fcd2}
\end{table}

\section{Practical Aspects: A protocol for running FDE-ET calculations}

\begin{figure}
\includegraphics[width=0.5\textheight]{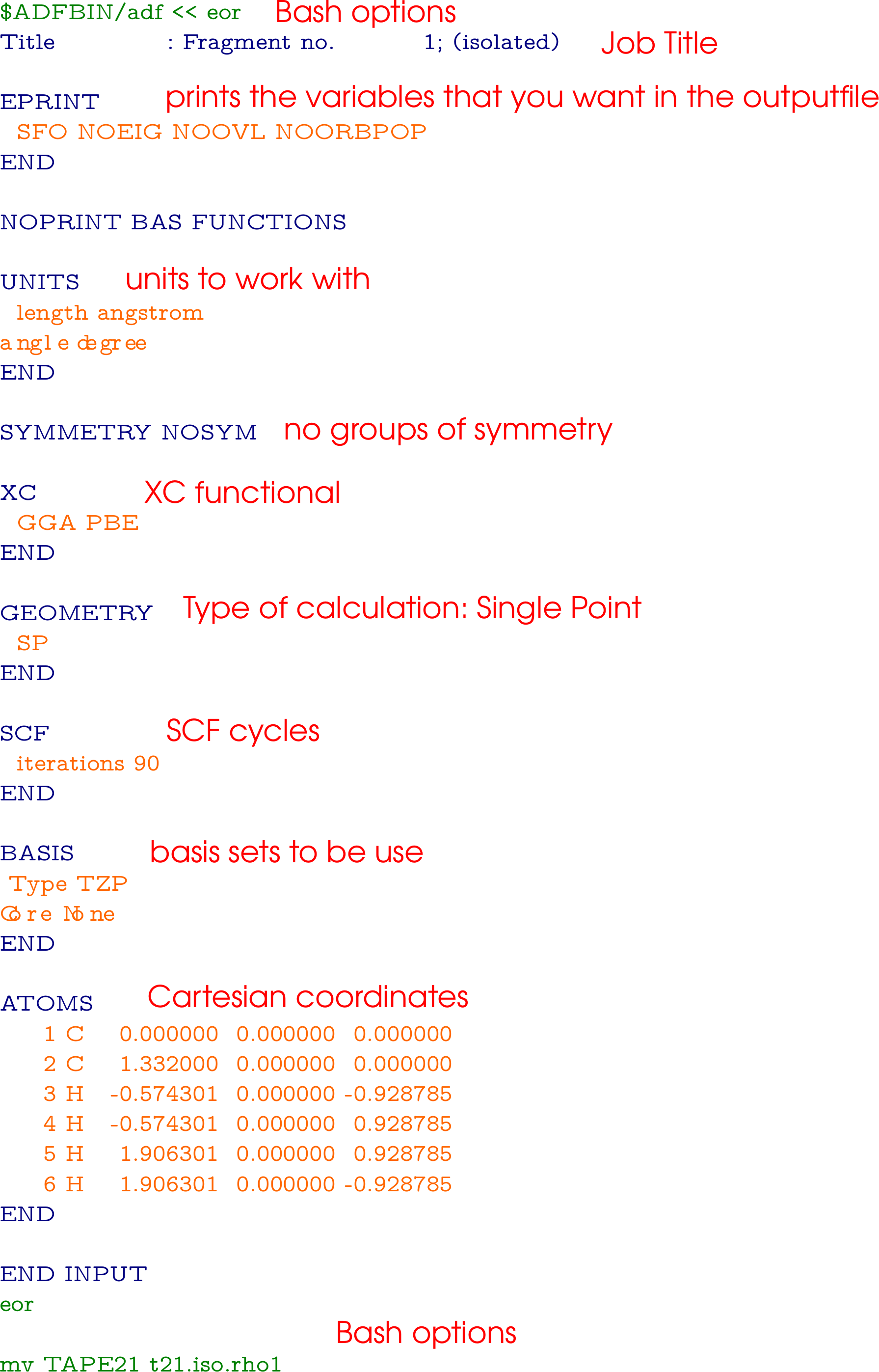}
\caption{\label{com1} Single point (SP) calculation input file for an isolated ethylene.}
\end{figure}

\begin{figure}
\includegraphics[width=0.35\textheight]{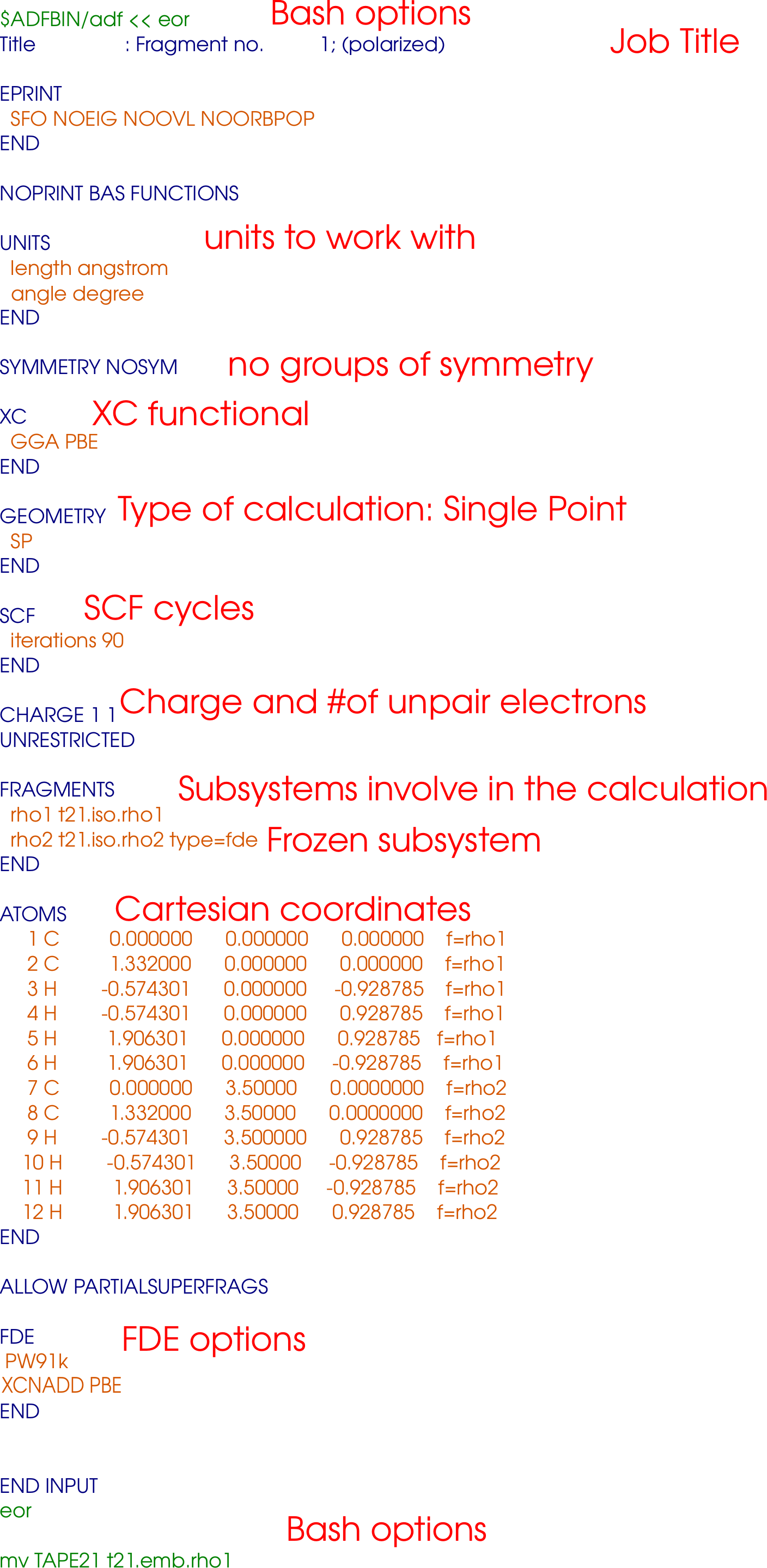}
\caption{\label{com2} Single point FDE calculation input file for ethylene dimer. Both fragments rho1 and rho2 come from two SP calculations for each isolated fragment.}
\end{figure}

\begin{figure}
\includegraphics[width=0.5\textheight]{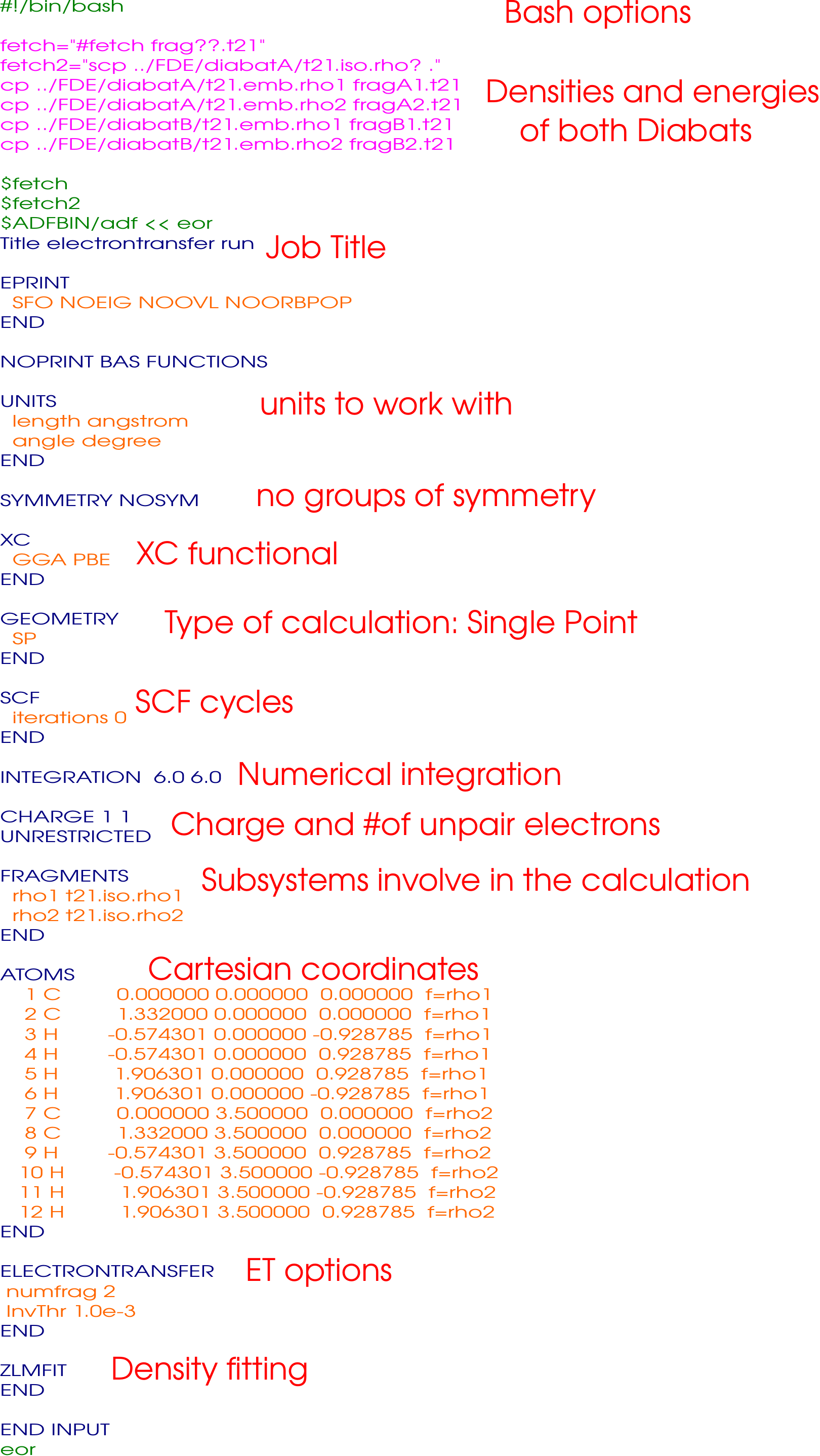}
\caption{\label{com3} Single point ET calculation input file for ethylene dimer. In pink there some bash-shell options in order to copy the information for each diabatic state calculated before with FDE. }
\end{figure}

In order to obtain the electronic coupling for a CT reaction using FDE-ET, three different single point (SP) calculations have to be performed. FDE-ET is available in ADF\cite{adf}. In Figures \ref{com1}, \ref{com2} and \ref{com3}, the input files corresponding to the FDE-ET methodology are described. First, a single point calculation for each isolated fragment present in the system is carried out. This gives the initial density and energy of each subsystem without any interaction between them. It is important to save the check point files (TAPE files in ADF), because they contain all fragment information needed in the subsequent calculations. Following all SP jobs for each isolated fragment, an FDE calculation is performed by taking into the account the whole supramolecular structure. So that, we create a diabatic state for each of the present subsystems. This is done by placing a charge different from neutrality in each subsystem, see Figure \ref{com2}. In this manner, two different directories are made: one in which an FDE calculation is carried out with subsystem 1 positively charged and one where subsystem 2 has the positive charge. In both cases, the SCF converges on the basis of subsystem DFT, thus, a series of three freeze--and--thaw procedure are done for each subsystem in each diabat. 

Once both isolated and embedded densities are obtained from the FDE calculations, a electrontransfer job is run whose purpose is to compute \eqs{cou}{coulow}. As in the FDE calculation, the information about the fragment is of paramount importance in this electrontransfer job. In figure \ref{com3}, the input file that calculates the diabatic energies and the electronic coupling between them is showed, in pink we can see that the check point files (t21.emb.rho* in the figure) corresponding to the embedded fragments in each diabat is copied directly to the ET directory, where the electrontransfer calculation is done. These files are renamed as fragA*.t21 for those ones from the diabat A (positive charge on the donor fragment) and fragB*.t21 for those that come from diabat B (positive charge on the acceptor fragment). It is worth mentioning that special care has to be taken in the management of the file names. As it is illustrated in Figure \ref{com3}, the fragments are numerated as 1 and 2, that means that the charge, departs from fragA1 while fragA2 is neutral and arrive to fragB2 while fragB1 becomes neutral. This is very important when the system is comprised of more than two fragments, and the charge is moving throughout all of them.

\section{Conclusions and future directions}
To conclude, we have presented our (fairly subjective) view of what tools are available nowadays to compute electronic couplings for charge transfer processes. We have surveyed in detail the FDE-ET method simply because we are among the developers of this method. Other methods based on DFT, and those that are best suited for being coupled with wavefunction based methods have also been discussed. The discussion also touches on the strengths and limitations of the various methods. 

When discussing the practical aspect of a coupling calculation, one must expose completely the methodology. We have done so for the FDE-ET method, and provided the reader with a step-by-step protocol on how to run such computations. This is important also for outsiders (such as experimentalists) as they can appreciate the kind of effort the theoretitians have to put in computing quantities relevant for the interpretation of the experiments.

We apologize in advance to those authors who have developed all those methods that we have omitted from this presentation. Admittedly, we provide here a subjective view of the field.

\section{Acknowledgements}
This work was funded by a grant from the National Science Foundation, Grant CBET-1438493.
We also acknowledge Dr. Eric Klein for bringing to our attention the problematics related to the electron transfer dynamics in the Sulfite Oxidase.

\bibliography{literatur}

\end{document}

%% file: macros.tex
\newcommand{\beq}{\begin{equation}}
\newcommand{\eeq}{\end{equation}}
\newcommand{\beqa}{\begin{eqnarray}}
\newcommand{\eeqa}{\end{eqnarray}}
\newcommand{\bfig}{\begin{figure}\begin{center}}
\newcommand{\efig}{\end{center}\end{figure}}
\newcommand{\btab}{\begin{table}\begin{center}}
\newcommand{\etab}{\end{center}\end{table}}

\newcommand{\eqn}[1]{\mbox{Eq.\hspace{1pt}(\ref{#1})}}
\newcommand{\eqs}[2]{\mbox{Eq.\hspace{1pt}(\ref{#1}--\ref{#2})}}

\newcommand{\mmfunc}[3]{#1_{#2} \left[ #3 \right] }

\newcommand{\pot}[1]{v_{\rm #1}}

\def\brp{{\mathbf{r}^{\prime}}}

\def\br{{\mathbf{r}}}

\def\bR{{\mathbf{R}}}

\def\d{{\mathrm{d}}}
\def\rhor{{\rho({\bf r})}}

\def\rhoi{{\rho_I}}

\def\rhoir{{\rho_I({\bf r})}}